\documentclass[runningheads,a4paper]{llncs}

\usepackage{amssymb}
\usepackage[dvips]{graphicx}

\usepackage[usenames,dvipsnames]{pstricks}
\usepackage{epsfig}
\usepackage{pst-grad} % For gradients
\usepackage{pst-plot} % For axes

\usepackage{url}
\usepackage[lined, boxruled, linesnumbered]{algorithm2e}

\begin{document}

\mainmatter  % start of an individual contribution

\title{A Memory versus Compression Ratio Trade--off  in {PPM} via Compressed Context Modeling}

% first the title is needed

% a short form should be given in case it is too long for the running head
\titlerunning{A Memory vs. Compression Ratio Trade--off  in {PPM} via CCM}

% the name(s) of the author(s) follow(s) next
%
% NB: Chinese authors should write their first names(s) in front of
% their surnames. This ensures that the names appear correctly in
% the running heads and the author index.
%
\author{M. O\v{g}uzhan K\"{u}lekci 
%\thanks{}
}
\authorrunning{K\"{u}lekci}

% the affiliations are given next; don't give your e-mail address
% unless you accept that it will be published
\institute{
T\"UB\.{I}TAK--B\.{I}LGEM--UEKAE\\
National Research Institute of Electronics \& Cryptology \\
41470 Gebze, Kocaeli, Turkey \\
%\vspace*{0.3cm}
\email{kulekci@uekae.tubitak.gov.tr}\\
}

%
% NB: a more complex sample for affiliations and the mapping to the
% corresponding authors can be found in the file "llncs.dem"
% (search for the string "\mainmatter" where a contribution starts).
% "llncs.dem" accompanies the document class "llncs.cls".
%

%\toctitle{Lecture Notes in Computer Science}
%\tocauthor{Authors' Instructions}
\maketitle

\sloppy

\begin{abstract}
Since its introduction prediction-by-partial-matching (PPM) has always been a de facto gold standard in lossless text
compression, where many variants improving the compression ratio and speed have been proposed. 
However, reducing the high space requirement of PPM schemes did not gain that much attention.
This study focuses on reducing the memory consumption of {PPM} via the recently proposed
\emph{compressed context modeling} (CCM) that uses the compressed representations of contexts in the
statistical model. 
Differently from the classical context definition  as the string  of the preceding characters at a particular position,
CCM considers context as the  amount of preceding information that is actually the bit-stream composed by compressing
the previous symbols.  
We observe that by using the {CCM}, the data structures, particularly the context trees, can be implemented in smaller
space, and present a trade--off between the compression ratio and the space requirement. The experiments conducted
showed that this trade--off is especially beneficial in low orders with a  $\approx 20$--$25$ percent gain
in memory by a sacrifice of up to $\approx 7$ percent loss in compression ratio. 
\end{abstract}

\section{Introduction}

Originating from the idea of predicting next symbol according to the statistics collected from the preceding symbols, 
prediction-by-partial-matching (PPM) has always been a de facto gold standard in lossless text compression since its
introduction. 
The major drawbacks of the scheme are huge memory consumption and slow speed, which prohibited its wide--spread usage in
practice for a long time, e.g.,  although the original scheme was proposed in $1984$ \cite{CW84},
it was not until $1990$ \cite{Moffat90} that the first practical implementation appeared as a consequence of limited
memory and computing power at the time of initial proposal. 

Many PPM variants \cite{Moffat90,WB91,HV94,CTW95,Bloom98,shkarin2002ppm}, which improved the compression ratio as
well as the speed, have been introduced during the last three decades. 
However, reducing the memory usage did not gain that much attention, and the advancement of technology has been assumed
the only source of progress in that direction. 

Today, it is true that we have plenty of memory in our computers, but also have many memory--hungry applications
demanding more space as well. 
The situation worsens on resource--limited environments such as the mobile phones or hand--held devices that are
surrounding the world in today's ubiquitous computing environment. 
When we consider the fact that the amount of data exchanged in wireless communication channels is increasing in an
unprecedented rate, where the users are billed according to the number of bytes they transmit, data
compression in mobile devices will apparently become more important in the very near future. 
Although {PPM} would be a strong option here, the statistical context modeling requiring significant run--time memory
may lack its practical impact on those resource--limited environments.

This study investigates ways to reduce the memory consumption of {PPM} via the recently proposed
\emph{compressed context modeling} (CCM) \cite{Kulekci11} that uses the compressed representations of contexts in the
statistical model. 
Differently from the classical context definition  as the string  of the preceding characters at a particular position,
the compressed context modeling considers context as the  amount of previous information that is actually the
bit-stream composed by compressing the preceding symbols.  
We observe that by using the {CCM}, the data structures, particularly the context trees, can be implemented in smaller
space. Based on this observation we present a compression--ratio/memory trade--off via CCM.  The experiments
conducted
showed that this trade--off is especially beneficial in low orders with a  $\approx 20$--$25$ percent gain
in memory by sacrificing up to $\approx 7$ percent loss in compression. 

As the outline of the paper, the PPM type compression scheme and several major improvements achieved previously 
are reviewed in section $2$. 
In section $3$, we introduce the main idea reducing the memory usage in PPM by the integration of  compressed
context modeling. 
The implementation of the proposed PPM$_{cc}$ is described in section $4$ that is followed by the experimental results
analyzing the trade--off between the compression ratio and the space requirement. 
We conclude with summary of the findings and future research directions.

\section{PPM and Previous Improvements}

In a Markovian sequence the appearance of a symbol at a specific position is assumed to be highly dependent on its
immediate predecessors. 
Cleary \& Witten \cite{CW84} introduced the prediction--by--partial--matching ({PPM}) compression scheme based on this
principle. 
The basic idea in PPM is to predict the next symbol based on its \emph{context}, which is defined as the
$k$--symbols preceding the current position. 
Assuming text $T$ of length $n$ as $T=t_1t_2 \ldots t_n$, the order--$k$ context of $t_i$, $ k < i \leq n$, is
$t_{i-1}t_{i-2} \ldots t_{i-k}$. 
The first step while encoding $t_i$ in PPM is to check whether its order--$k$ context has ever been followed by $t_i$
previously.
If the pattern $t_{i-k}t_{i-k+1} \ldots t_{i-1}t_i$ has been observed in $t_1t_2 \ldots t_{i-1}$, the probability
$P(t_i \;|\; t_{i-k}t_{i-k+1} \ldots t_{i-1})$ is sent to the entropy encoder. 
Otherwise, an escape symbol is emitted with the probability $P( escape \;|\; t_{i-k}t_{i-k+1} \ldots t_{i-1})$, which 
is computed according to the zero-frequency handling of the used statistical model, and the length of the context is
shortened usually by decrementing $k$ by $1$.
The same procedure is repeated with the shortened context until a non-zero probability is obtained for $t_i$ 
or the context length becomes zero. If the probability of $t_i$ is still missing in the order-$0$ model, which means
$t_i$ is a novel symbol appearing first at position $i$, the encoding of the actual raw value is
performed after the emission of the escape. 

The compression performance of PPM improves with increasing length of the context. 
However, after some certain point the compression ratio begins deteriorating. 
On the Calgary corpus, it is reported that after order-$6$ the compression becomes going downhill \cite{Sal04}. 
That is due to the sparseness of the statistics, which accelerates rapidly with the increasing length of the context, 
 and causes long chains of escape symbols. 
Therefore, much of the attention to improve PPM has been paid in the direction of computing the probability of a 
zero-frequency item, which is also a fundamental problem in statistics \cite{WB91,CT95,wood2011}. 
In that sense, different approaches of calculating the escape and symbol probabilities led to PPM variants such as PPMC
\cite{Moffat90}, PPMP and PPMX \cite{WB91}, and FastPPM \cite{HV94}. The PPMII \cite{shkarin2001,shkarin2002ppm} scheme
of Shkarin introduced an information inheritance mechanism  that the missing probabilities in long contexts are 
estimated from their shorter subsequences. 
More recently,  a lossless compression scheme based on the sequence memoizer \cite{gasthaus2010} has been introduced
which improves the compression ratio by enhanced symbol and escape probability estimations.

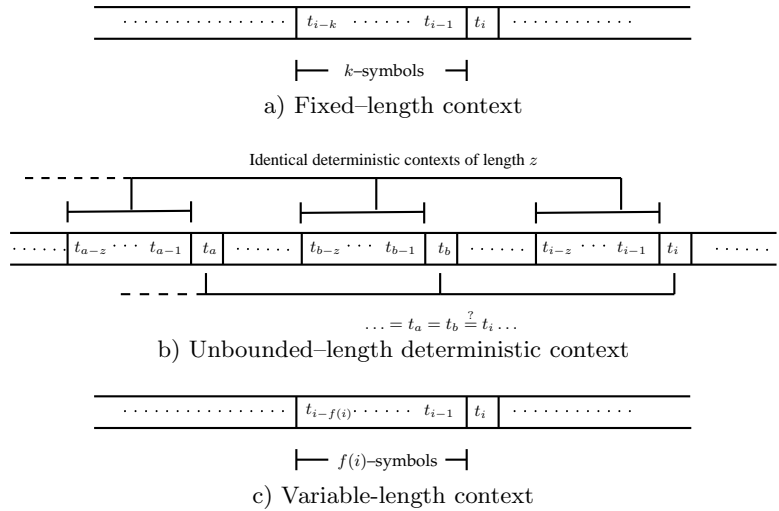
\begin{figure}
\begin{center}
 \begin{tabular}{c}
\scalebox{.7} % Change this value to rescale the drawing.
{
\begin{pspicture}(0,-0.7475)(11.24,0.7275)
\usefont{T1}{ppl}{m}{n}
\rput(7.298281,0.4175){$t_i$ }
\usefont{T1}{ppl}{m}{n}
\rput(4.288906,0.4175){$t_{i-k}$}
\usefont{T1}{ppl}{m}{n}
\rput(6.483125,0.4175){$t_{i-1}$}
\psline[linewidth=0.04cm,linestyle=dotted,dotsep=0.16cm](4.88,0.4275)(5.92,0.4275)
\psline[linewidth=0.04cm](0.0,0.7075)(11.2,0.7075)
\psline[linewidth=0.04cm](0.0,0.1075)(11.22,0.1075)
\psline[linewidth=0.04cm](3.8,0.7075)(3.8,0.1075)
\psline[linewidth=0.04cm](7.0,0.7075)(7.0,0.1075)
\psline[linewidth=0.04cm](7.6,0.7075)(7.6,0.1075)
\psline[linewidth=0.04cm,linestyle=dotted,dotsep=0.16cm](0.56,0.4275)(3.56,0.4275)
\psline[linewidth=0.04cm,linestyle=dotted,dotsep=0.16cm](7.88,0.4275)(10.08,0.4275)
\usefont{T1}{ppl}{m}{n}
\rput(5.4485936,-0.5225){$k$--symbols}
\psline[linewidth=0.04cm](3.8,-0.4925)(4.4,-0.4925)
\psline[linewidth=0.04cm](6.6,-0.4925)(7.0,-0.4925)
\psline[linewidth=0.04cm](7.0,-0.2925)(7.0,-0.6925)
\psline[linewidth=0.04cm](3.8,-0.2925)(3.8,-0.6925)
\end{pspicture} 
}
\\
  a) Fixed--length context \\
\\
\scalebox{.7} % Change this value to rescale the drawing.
{
\begin{pspicture}(0,-1.6792188)(14.44,1.6792188)
\usefont{T1}{ptm}{m}{n}
\rput(3.7790625,-0.16421875){$t_a$}
\usefont{T1}{ptm}{m}{n}
\rput(1.541875,-0.16421875){$t_{a-z}$}
\usefont{T1}{ptm}{m}{n}
\rput(2.953125,-0.16421875){$t_{a-1}$}
\psline[linewidth=0.04cm,linestyle=dotted,dotsep=0.16cm](1.96,-0.11421875)(2.4,-0.11421875)
\psline[linewidth=0.04cm](0.02,0.12578125)(14.42,0.12578125)
\psline[linewidth=0.04cm](0.02,-0.47421876)(14.42,-0.47421876)
\psline[linewidth=0.04cm](1.1,0.12578125)(1.1,-0.47421876)
\psline[linewidth=0.04cm,linestyle=dotted,dotsep=0.16cm](0.08,-0.17421874)(1.06,-0.19421875)
\psline[linewidth=0.04cm](3.42,0.12578125)(3.42,-0.47421876)
\psline[linewidth=0.04cm](4.02,0.12578125)(4.02,-0.47421876)
\usefont{T1}{ptm}{m}{n}
\rput(8.189531,-0.16421875){$t_b$}
\usefont{T1}{ptm}{m}{n}
\rput(5.951875,-0.16421875){$t_{b-z}$}
\usefont{T1}{ptm}{m}{n}
\rput(7.363125,-0.16421875){$t_{b-1}$}
\psline[linewidth=0.04cm,linestyle=dotted,dotsep=0.16cm](6.36,-0.11421875)(6.8,-0.11421875)
\psline[linewidth=0.04cm](5.5,0.12578125)(5.5,-0.47421876)
\psline[linewidth=0.04cm](7.82,0.12578125)(7.82,-0.47421876)
\psline[linewidth=0.04cm](8.42,0.12578125)(8.42,-0.47421876)
\usefont{T1}{ptm}{m}{n}
\rput(12.518281,-0.16421875){$t_i$ }
\usefont{T1}{ptm}{m}{n}
\rput(10.308907,-0.16421875){$t_{i-z}$}
\usefont{T1}{ptm}{m}{n}
\rput(11.703125,-0.16421875){$t_{i-1}$}
\psline[linewidth=0.04cm,linestyle=dotted,dotsep=0.16cm](10.76,-0.11421875)(11.2,-0.11421875)
\psline[linewidth=0.04cm](9.9,0.12578125)(9.9,-0.47421876)
\psline[linewidth=0.04cm](12.22,0.12578125)(12.22,-0.47421876)
\psline[linewidth=0.04cm](12.82,0.12578125)(12.82,-0.47421876)
\psline[linewidth=0.04cm,linestyle=dotted,dotsep=0.16cm](4.28,-0.17421874)(5.26,-0.19421875)
\psline[linewidth=0.04cm,linestyle=dotted,dotsep=0.16cm](8.68,-0.17421874)(9.66,-0.19421875)
\psline[linewidth=0.04cm,linestyle=dotted,dotsep=0.16cm](13.28,-0.17421874)(14.26,-0.19421875)
\psline[linewidth=0.04cm](1.1,0.72578126)(1.1,0.32578126)
\psline[linewidth=0.04cm](3.42,0.72578126)(3.42,0.32578126)
\psline[linewidth=0.04cm](1.1,0.5257813)(3.42,0.54578125)
\psline[linewidth=0.04cm](5.48,0.7057812)(5.48,0.30578125)
\psline[linewidth=0.04cm](7.8,0.7057812)(7.8,0.30578125)
\psline[linewidth=0.04cm](5.48,0.50578123)(7.8,0.5257813)
\psline[linewidth=0.04cm](9.9,0.7057812)(9.9,0.30578125)
\psline[linewidth=0.04cm](12.22,0.7057812)(12.22,0.30578125)
\psline[linewidth=0.04cm](9.9,0.50578123)(12.22,0.5257813)
\psline[linewidth=0.04cm](2.3,1.1657813)(2.3,0.56578124)
\psline[linewidth=0.04cm](6.9,1.1657813)(6.9,0.56578124)
\psline[linewidth=0.04cm](11.5,1.1657813)(11.5,0.56578124)
\psline[linewidth=0.04cm](2.3,1.1657813)(11.5,1.1657813)
\psline[linewidth=0.04cm,linestyle=dashed,dash=0.16cm 0.16cm](0.3,1.1657813)(2.3,1.1657813)
\usefont{T1}{ptm}{m}{n}
\rput(7.2290626,1.4757812){Identical deterministic contexts of length $z$}
\psline[linewidth=0.04cm](3.7,-0.63421875)(3.7,-1.0342188)
\psline[linewidth=0.04cm](8.1,-0.63421875)(8.1,-1.0342188)
\psline[linewidth=0.04cm](3.7,-1.0342188)(8.1,-1.0342188)
\psline[linewidth=0.04cm,linestyle=dashed,dash=0.16cm 0.16cm](2.1,-1.0342188)(3.7,-1.0342188)
\usefont{T1}{ptm}{m}{n}
\rput(8.158594,-1.5242188){$ \ldots = t_a = t_b \stackrel{?}{=} t_i \ldots$ }
\psline[linewidth=0.04cm](8.1,-1.0342188)(12.5,-1.0342188)
\psline[linewidth=0.04cm](12.5,-1.0342188)(12.5,-0.63421875)
\end{pspicture} 
}
\\
b) Unbounded--length deterministic context \\
\\
\scalebox{.7} % Change this value to rescale the drawing.
{
\begin{pspicture}(0,-0.7575)(11.24,0.7375)
\usefont{T1}{ppl}{m}{n}
\rput(7.298281,0.4275){$t_i$ }
\usefont{T1}{ppl}{m}{n}
\rput(4.4198437,0.4275){$t_{i-f(i)}$}
\usefont{T1}{ppl}{m}{n}
\rput(6.483125,0.4275){$t_{i-1}$}
\psline[linewidth=0.04cm,linestyle=dotted,dotsep=0.16cm](4.88,0.4375)(5.92,0.4375)
\psline[linewidth=0.04cm](0.0,0.7175)(11.2,0.7175)
\psline[linewidth=0.04cm](0.0,0.1175)(11.22,0.1175)
\psline[linewidth=0.04cm](3.8,0.7175)(3.8,0.1175)
\psline[linewidth=0.04cm](7.0,0.7175)(7.0,0.1175)
\psline[linewidth=0.04cm](7.6,0.7175)(7.6,0.1175)
\psline[linewidth=0.04cm,linestyle=dotted,dotsep=0.16cm](0.56,0.4375)(3.56,0.4375)
\psline[linewidth=0.04cm,linestyle=dotted,dotsep=0.16cm](7.88,0.4375)(10.08,0.4375)
\usefont{T1}{ppl}{m}{n}
\rput(5.5109377,-0.5325){$f(i)$--symbols}
\psline[linewidth=0.04cm](3.8,-0.4825)(4.4,-0.4825)
\psline[linewidth=0.04cm](6.6,-0.4825)(7.0,-0.4825)
\psline[linewidth=0.04cm](7.0,-0.2825)(7.0,-0.6825)
\psline[linewidth=0.04cm](3.8,-0.2825)(3.8,-0.6825)
\end{pspicture} 
}
\\
c) Variable-length context\\
 \end{tabular}
\caption{Context modeling techniques}
\label{fig:contextmodelingtechniques}
\end{center}
\end{figure}

Another direction of research has been to apply different context modeling \cite{Bell89,Mahoney05} approaches that are
depicted in Figure \ref{fig:contextmodelingtechniques}. 
In early studies \cite{CW84,Moffat90,HV94} the context was assumed to be a fixed--size  window of preceding symbols at a
particular position. 
As oppose to using this bounded-length windows, the \emph{deterministic context} \cite{CT97} was proposed with PPM*. 
While encoding the $i^{th}$ character $t_i$, the PPM* first investigates whether there exists a string
$t_{i-z} \ldots t_{i-1}$ of arbitrary length $z$ ($z>i$), which is always followed by the same symbol $t_i$. 
In case such a deterministic context exists, the encoding can be achieved very effectively.  
Otherwise, the PPM* switches to the classic bounded-length model. 
Instead of switching to a fixed--order model, the PPMZ \cite{Bloom98} scheme following the PPM* proposed
using \emph{local-order-estimation} ({LOE}) that decides on the order of the context at a position according to a
heuristic function. 
Therefore, the variable-order context idea had first appeared with the PPMZ. 
Another variant of PPM that uses variable-order modeling is the PPMVC \cite{skibinski2004} that also benefits from
PPMII idea.
Note that both the unbounded--length deterministic context investigations as well as the variable-order calculations
require additional computational resources with an increased memory consumption. 

\section{Preserving Memory via Compressed Context Modeling}

On a given text $T=t_1t_2 \ldots t_n$ of length $n$, where the individual characters are drawn from alphabet
$\Sigma$, the order--$k$ compressed context of $t_i$ is the first $k$--bits of the preceding \emph{information} 
that is computed via compressing the string $t_{i-\ell} \ldots t_{i-1}$ for sufficiently large $\ell$,
$1 \leq \ell < i$. 
Assuming $\mathcal{C}$ is a proper compression function, we search for the smallest $\ell$ such that 
the length of the bit-stream formed by $\mathcal{C}(t_{i-1}t_{i-2} \ldots t_{i-\ell})$ is larger than or equal to $k$. 

\begin{figure}
\begin{center}
\begin{tabular}{c}
\scalebox{.5} % Change this value to rescale the drawing.
{
\begin{pspicture}(0,-2.63)(19.05,2.62)
\psframe[linewidth=0.04,dimen=outer](11.42,2.6)(7.02,2.0)
\psline[linewidth=0.04cm](7.82,2.6)(7.82,2.0)
\usefont{T1}{ppl}{m}{n}
\rput(7.397969,2.31){$c_1$}
\psline[linewidth=0.04cm](8.62,2.6)(8.62,2.0)
\psline[linewidth=0.04cm](10.62,2.6)(10.62,2.0)
\usefont{T1}{ppl}{m}{n}
\rput(8.196563,2.31){$c_2$}
\usefont{T1}{ppl}{m}{n}
\rput(11.002188,2.31){$c_{|\Sigma|}$}
\psframe[linewidth=0.04,dimen=outer](11.42,2.0)(7.02,1.4)
\psline[linewidth=0.04cm](7.82,2.0)(7.82,1.4)
\usefont{T1}{ppl}{m}{n}
\rput(7.4115624,1.71){$p_1$}
\psline[linewidth=0.04cm](8.62,2.0)(8.62,1.4)
\psline[linewidth=0.04cm](10.62,2.0)(10.62,1.4)
\usefont{T1}{ppl}{m}{n}
\rput(8.210156,1.71){$p_2$}
\usefont{T1}{ppl}{m}{n}
\rput(11.015781,1.71){$p_{|\Sigma|}$}
\psline[linewidth=0.04cm](9.42,2.6)(9.42,1.4)
\usefont{T1}{ppl}{m}{n}
\rput(9.01375,1.71){$p_3$}
\usefont{T1}{ppl}{m}{n}
\rput(9.000156,2.31){$c_3$}
\psline[linewidth=0.04cm](7.42,1.4)(3.22,-0.2)
\psframe[linewidth=0.04,dimen=outer](4.42,-0.2)(0.02,-0.8)
\psline[linewidth=0.04cm](0.82,-0.2)(0.82,-0.8)
\usefont{T1}{ppl}{m}{n}
\rput(0.39796874,-0.49){$c_1$}
\psline[linewidth=0.04cm](1.62,-0.2)(1.62,-0.8)
\psline[linewidth=0.04cm](3.62,-0.2)(3.62,-0.8)
\usefont{T1}{ppl}{m}{n}
\rput(1.1965625,-0.49){$c_2$}
\usefont{T1}{ppl}{m}{n}
\rput(4.0021877,-0.49){$c_{|\Sigma|}$}
\psframe[linewidth=0.04,dimen=outer](4.42,-0.8)(0.02,-1.4)
\psline[linewidth=0.04cm](0.82,-0.8)(0.82,-1.4)
\usefont{T1}{ppl}{m}{n}
\rput(0.4115625,-1.09){$p_1$}
\psline[linewidth=0.04cm](1.62,-0.8)(1.62,-1.4)
\psline[linewidth=0.04cm](3.62,-0.8)(3.62,-1.4)
\usefont{T1}{ppl}{m}{n}
\rput(1.2101562,-1.09){$p_2$}
\usefont{T1}{ppl}{m}{n}
\rput(4.0157814,-1.09){$p_{|\Sigma|}$}
\psline[linewidth=0.04cm](2.42,-0.2)(2.42,-1.4)
\usefont{T1}{ppl}{m}{n}
\rput(2.01375,-1.09){$p_3$}
\usefont{T1}{ppl}{m}{n}
\rput(2.0001562,-0.49){$c_3$}
\psline[linewidth=0.04cm](0.02,-0.8)(4.42,-0.8)
\psline[linewidth=0.04cm](0.42,-1.4)(0.02,-2.0)
\psline[linewidth=0.04cm](1.22,-1.4)(1.22,-2.0)
\psline[linewidth=0.04cm](2.02,-1.4)(2.42,-2.0)
\psline[linewidth=0.04cm](4.02,-1.4)(4.42,-2.0)
\psline[linewidth=0.04cm](9.02,1.4)(9.82,-0.2)
\psline[linewidth=0.04cm](8.22,1.4)(5.62,-0.2)
\psline[linewidth=0.04cm](11.02,1.4)(15.82,-0.2)
\psline[linewidth=0.06cm,linestyle=dashed,dash=0.16cm 0.16cm](11.62,-0.8)(14.22,-0.8)
\psline[linewidth=0.04cm](5.22,-0.2)(6.02,-0.2)
\psline[linewidth=0.04cm](5.42,-0.4)(5.82,-0.4)
\psdots[dotsize=0.12](5.62,-0.6)
\psline[linewidth=0.06cm,linestyle=dashed,dash=0.16cm 0.16cm](0.22,-2.6)(19.02,-2.6)
\psline[linewidth=0.04cm](7.02,2.0)(11.42,2.0)
\psframe[linewidth=0.04,dimen=outer](11.22,-0.2)(6.82,-0.8)
\psline[linewidth=0.04cm](7.62,-0.2)(7.62,-0.8)
\usefont{T1}{ppl}{m}{n}
\rput(7.197969,-0.49){$c_1$}
\psline[linewidth=0.04cm](8.42,-0.2)(8.42,-0.8)
\psline[linewidth=0.04cm](10.42,-0.2)(10.42,-0.8)
\usefont{T1}{ppl}{m}{n}
\rput(7.9965625,-0.49){$c_2$}
\usefont{T1}{ppl}{m}{n}
\rput(10.802188,-0.49){$c_{|\Sigma|}$}
\psframe[linewidth=0.04,dimen=outer](11.22,-0.8)(6.82,-1.4)
\psline[linewidth=0.04cm](7.62,-0.8)(7.62,-1.4)
\usefont{T1}{ppl}{m}{n}
\rput(7.2115626,-1.09){$p_1$}
\psline[linewidth=0.04cm](8.42,-0.8)(8.42,-1.4)
\psline[linewidth=0.04cm](10.42,-0.8)(10.42,-1.4)
\usefont{T1}{ppl}{m}{n}
\rput(8.010157,-1.09){$p_2$}
\usefont{T1}{ppl}{m}{n}
\rput(10.815782,-1.09){$p_{|\Sigma|}$}
\psline[linewidth=0.04cm](9.22,-0.2)(9.22,-1.4)
\usefont{T1}{ppl}{m}{n}
\rput(8.81375,-1.09){$p_3$}
\usefont{T1}{ppl}{m}{n}
\rput(8.800157,-0.49){$c_3$}
\psline[linewidth=0.04cm](6.82,-0.8)(11.22,-0.8)
\psline[linewidth=0.04cm](7.22,-1.4)(6.82,-2.0)
\psline[linewidth=0.04cm](8.02,-1.4)(8.02,-2.0)
\psline[linewidth=0.04cm](8.82,-1.4)(9.22,-2.0)
\psline[linewidth=0.04cm](10.82,-1.4)(11.22,-2.0)
\psframe[linewidth=0.04,dimen=outer](18.82,-0.2)(14.42,-0.8)
\psline[linewidth=0.04cm](15.22,-0.2)(15.22,-0.8)
\usefont{T1}{ppl}{m}{n}
\rput(14.797969,-0.49){$c_1$}
\psline[linewidth=0.04cm](16.02,-0.2)(16.02,-0.8)
\psline[linewidth=0.04cm](18.02,-0.2)(18.02,-0.8)
\usefont{T1}{ppl}{m}{n}
\rput(15.596562,-0.49){$c_2$}
\usefont{T1}{ppl}{m}{n}
\rput(18.402187,-0.49){$c_{|\Sigma|}$}
\psframe[linewidth=0.04,dimen=outer](18.82,-0.8)(14.42,-1.4)
\psline[linewidth=0.04cm](15.22,-0.8)(15.22,-1.4)
\usefont{T1}{ppl}{m}{n}
\rput(14.811563,-1.09){$p_1$}
\psline[linewidth=0.04cm](16.02,-0.8)(16.02,-1.4)
\psline[linewidth=0.04cm](18.02,-0.8)(18.02,-1.4)
\usefont{T1}{ppl}{m}{n}
\rput(15.610156,-1.09){$p_2$}
\usefont{T1}{ppl}{m}{n}
\rput(18.415781,-1.09){$p_{|\Sigma|}$}
\psline[linewidth=0.04cm](16.82,-0.2)(16.82,-1.4)
\usefont{T1}{ppl}{m}{n}
\rput(16.41375,-1.09){$p_3$}
\usefont{T1}{ppl}{m}{n}
\rput(16.400156,-0.49){$c_3$}
\psline[linewidth=0.04cm](14.42,-0.8)(18.82,-0.8)
\psline[linewidth=0.04cm](14.82,-1.4)(14.42,-2.0)
\psline[linewidth=0.04cm](15.62,-1.4)(15.62,-2.0)
\psline[linewidth=0.04cm](16.42,-1.4)(16.82,-2.0)
\psline[linewidth=0.04cm](18.42,-1.4)(18.82,-2.0)
\usefont{T1}{ptm}{m}{n}
\rput(5.0179687,0.91){$\epsilon_1$}
\usefont{T1}{ptm}{m}{n}
\rput(7.4165626,0.51){$\epsilon_2$}
\usefont{T1}{ptm}{m}{n}
\rput(10.020156,0.51){$\epsilon_3$}
\usefont{T1}{ptm}{m}{n}
\rput(14.222187,0.71){$\epsilon_{|\Sigma|}$}
\psline[linewidth=0.06cm,linestyle=dashed,dash=0.16cm 0.16cm](11.02,0.4)(13.02,0.4)
\usefont{T1}{ptm}{m}{n}
\rput(0.61796874,-1.69){$\epsilon_1$}
\usefont{T1}{ptm}{m}{n}
\rput(1.6165625,-1.69){$\epsilon_2$}
\usefont{T1}{ptm}{m}{n}
\rput(2.6201563,-1.69){$\epsilon_3$}
\usefont{T1}{ptm}{m}{n}
\rput(4.6221876,-1.69){$\epsilon_{|\Sigma|}$}
\psline[linewidth=0.06cm,linestyle=dashed,dash=0.16cm 0.16cm](3.02,-1.8)(3.82,-1.8)
\usefont{T1}{ptm}{m}{n}
\rput(7.4179688,-1.69){$\epsilon_1$}
\usefont{T1}{ptm}{m}{n}
\rput(8.416562,-1.69){$\epsilon_2$}
\usefont{T1}{ptm}{m}{n}
\rput(9.4201565,-1.69){$\epsilon_3$}
\usefont{T1}{ptm}{m}{n}
\rput(11.422188,-1.69){$\epsilon_{|\Sigma|}$}
\psline[linewidth=0.06cm,linestyle=dashed,dash=0.16cm 0.16cm](9.82,-1.8)(10.62,-1.8)
\usefont{T1}{ptm}{m}{n}
\rput(15.017969,-1.69){$\epsilon_1$}
\usefont{T1}{ptm}{m}{n}
\rput(16.016563,-1.69){$\epsilon_2$}
\usefont{T1}{ptm}{m}{n}
\rput(17.020157,-1.69){$\epsilon_3$}
\usefont{T1}{ptm}{m}{n}
\rput(19.022188,-1.69){$\epsilon_{|\Sigma|}$}
\psline[linewidth=0.06cm,linestyle=dashed,dash=0.16cm 0.16cm](17.42,-1.8)(18.22,-1.8)
\end{pspicture} 
} \\
\\
a) Classical Context Trie \\

  \\
\\

\scalebox{.5} % Change this value to rescale the drawing.
{
\begin{pspicture}(0,-3.13)(11.23,3.12)
\psframe[linewidth=0.04,dimen=outer](7.8,3.1)(3.4,2.5)
\psline[linewidth=0.04cm](4.2,3.1)(4.2,2.5)
\usefont{T1}{ppl}{m}{n}
\rput(3.7779686,2.81){$c_1$}
\psline[linewidth=0.04cm](5.0,3.1)(5.0,2.5)
\psline[linewidth=0.04cm](7.0,3.1)(7.0,2.5)
\usefont{T1}{ppl}{m}{n}
\rput(4.5765624,2.81){$c_2$}
\usefont{T1}{ppl}{m}{n}
\rput(7.3821874,2.81){$c_{|\Sigma|}$}
\psframe[linewidth=0.04,dimen=outer](6.4,2.5)(4.8,1.9)
\psline[linewidth=0.04cm](5.6,2.5)(5.6,1.9)
\usefont{T1}{ppl}{m}{n}
\rput(5.1960936,2.21){$p_0$}
\usefont{T1}{ppl}{m}{n}
\rput(5.9915624,2.21){$p_1$}
\psline[linewidth=0.04cm](5.0,2.5)(5.0,2.5)
\psline[linewidth=0.06cm,linestyle=dashed,dash=0.16cm 0.16cm](5.44,2.8)(6.64,2.8)
\psline[linewidth=0.04cm](4.8,2.5)(6.4,2.5)
\psframe[linewidth=0.04,dimen=outer](10.6,0.7)(6.2,0.1)
\psline[linewidth=0.04cm](7.0,0.7)(7.0,0.1)
\usefont{T1}{ppl}{m}{n}
\rput(6.5779686,0.41){$c_1$}
\psline[linewidth=0.04cm](7.8,0.7)(7.8,0.1)
\psline[linewidth=0.04cm](9.8,0.7)(9.8,0.1)
\usefont{T1}{ppl}{m}{n}
\rput(7.3765626,0.41){$c_2$}
\usefont{T1}{ppl}{m}{n}
\rput(10.182187,0.41){$c_{|\Sigma|}$}
\psframe[linewidth=0.04,dimen=outer](9.2,0.1)(7.6,-0.5)
\psline[linewidth=0.04cm](8.4,0.1)(8.4,-0.5)
\usefont{T1}{ppl}{m}{n}
\rput(7.9960938,-0.19){$p_0$}
\usefont{T1}{ppl}{m}{n}
\rput(8.791562,-0.19){$p_1$}
\psline[linewidth=0.04cm](7.8,0.1)(7.8,0.1)
\psline[linewidth=0.06cm,linestyle=dashed,dash=0.16cm 0.16cm](8.24,0.4)(9.44,0.4)
\psline[linewidth=0.04cm](7.6,0.1)(9.2,0.1)
\psframe[linewidth=0.04,dimen=outer](5.0,0.7)(0.6,0.1)
\psline[linewidth=0.04cm](1.4,0.7)(1.4,0.1)
\usefont{T1}{ppl}{m}{n}
\rput(0.97796875,0.41){$c_1$}
\psline[linewidth=0.04cm](2.2,0.7)(2.2,0.1)
\psline[linewidth=0.04cm](4.2,0.7)(4.2,0.1)
\usefont{T1}{ppl}{m}{n}
\rput(1.7765625,0.41){$c_2$}
\usefont{T1}{ppl}{m}{n}
\rput(4.5821877,0.41){$c_{|\Sigma|}$}
\psframe[linewidth=0.04,dimen=outer](3.6,0.1)(2.0,-0.5)
\psline[linewidth=0.04cm](2.8,0.1)(2.8,-0.5)
\usefont{T1}{ppl}{m}{n}
\rput(2.3960938,-0.19){$p_0$}
\usefont{T1}{ppl}{m}{n}
\rput(3.1915624,-0.19){$p_1$}
\psline[linewidth=0.04cm](2.2,0.1)(2.2,0.1)
\psline[linewidth=0.06cm,linestyle=dashed,dash=0.16cm 0.16cm](2.64,0.4)(3.84,0.4)
\psline[linewidth=0.04cm](2.0,0.1)(3.6,0.1)
\psline[linewidth=0.04cm](5.2,1.9)(2.8,0.7)
\psline[linewidth=0.04cm](6.0,1.9)(8.6,0.7)
\psline[linewidth=0.04cm](2.2,-0.5)(1.2,-1.9)
\psline[linewidth=0.04cm](3.2,-0.5)(4.0,-1.9)
\psline[linewidth=0.04cm](8.0,-0.5)(7.0,-1.9)
\psline[linewidth=0.04cm](8.8,-0.5)(9.8,-1.9)
\psline[linewidth=0.04cm](6.6,-1.9)(7.4,-1.9)
\psline[linewidth=0.04cm](6.8,-2.1)(7.2,-2.1)
\psdots[dotsize=0.12](7.0,-2.3)
\psline[linewidth=0.06cm,linestyle=dashed,dash=0.16cm 0.16cm](0.0,-3.1)(11.2,-3.1)
\usefont{T1}{ptm}{m}{n}
\rput(3.690625,1.41){0}
\usefont{T1}{ptm}{m}{n}
\rput(7.678125,1.41){1}
\usefont{T1}{ptm}{m}{n}
\rput(1.290625,-1.19){0}
\usefont{T1}{ptm}{m}{n}
\rput(3.878125,-1.19){1}
\usefont{T1}{ptm}{m}{n}
\rput(7.090625,-1.19){0}
\usefont{T1}{ptm}{m}{n}
\rput(9.678125,-1.19){1}
\end{pspicture} 
} \\
\\
b) Compressed Context Trie 
\end{tabular}
\end{center}
\caption{Context trie structures in classical and compressed context modeling.}
\label{fig:ContextTrie}
\end{figure}

In this work, we use a standard $0^{th}$--order Huffman coding as the $\mathcal{C}$ compression function. 
%Although this is not a very competitive way of representing the information content of a given string, it is preferred
%for simplicity and ease--of--use in this  proof--of--concept study of CCM integration. 
Obviously, the Huffman code table should also be attached to the final compressed file, which brings an overhead. 
Since using a higher order Huffman will enlarge that overhead, the choice of $0^{th}$--order is supposed to be a better
fit especially in case of small files as in the Large Calgary corpus. 
One may argue that it is possible to integrate a dynamic Huffman compression as well, where we do not need to carry any
additional information and also do not need to perform an initial scan over the file to calculate the Huffman tree.
Since such dynamic approaches work better on longer files, they are not included in this study as we basically aim to
see whether CCM can achieve a space preservation in its simplest settings. Future studies are expected to investigate
more options in that sense. 

The data structures used in context modeling are either the hash tables \cite{RT95} or the context tries. 
Since slow speed is one of the major problems in PPM, most of the implementations in practice prefer context trie data
structures to achieve fast processing.   
Each node in the trie simply holds the frequency counts of the symbols and pointers to next nodes. 
Figure \ref{fig:ContextTrie}--\emph{a} simply exhibits such a trie in classical setting. 
Observe that the number of pointers and the frequency counts are in order of the alphabet size of the input data,
where a standard general purpose implementation should consider the $256$--bytes ASCII table in practice. 

On the other side, the trie structure that is used when one prefers CCM is sketched in Figure 
\ref{fig:ContextTrie}--\emph{b}. 
As oppose to the classical $|\Sigma|$-ary context trees, CCM--integrated PPM implementation uses binary trees regardless
of the alphabet size as the compressed context by definition is a bit-stream of just $0$s and $1$s. 
Therefore, at each node we do not need to reserve $|\Sigma|$ pointers for every character of the alphabet, but only two
pointers for the next bit.

The gain in space via CCM reported in this study actually stems from that reduction of the size of the nodes in the
context trie. 
Note that applying some dynamic data allocation tricks to use less space in the trie structures are possible. However,
such attempts slow down the execution time, and hence, are contrary to the goal of achieving fast compression 
(decompression).
Thus, we neglect these kind of programming practices.

\section{Implementation Details}

We have implemented the basic PPM (following the Moffat's study \cite{Moffat90}), and its proposed variant PPM$_{cc}$
with the compressed context modeling. 
While computing the escape and symbol probabilities in both implementations, we used the method proposed by
Howard\&Vitter \cite{HV92} as this technique consistently achieved better than the other alternatives. 
For entropy coding/decoding, we preferred the \texttt{FastAC}\footnote{Available at
\url{http://www.cipr.rpi.edu/~said/FastAC.html}} arithmetic coder of Said \cite{said2004introduction}. 

The PPM$_{cc}$ replaces the classical context modeling in the basic PPM with the CCM. 
In order to be able to compute the compressed context of a position via $0^{th}$--order Huffman coding, we first make
an initial pass over the input file and generate the corresponding Huffman tree. 
The probability of the next symbol is estimated according to the first $k$--bits of the preceding compressed context
instead of the most immediate symbols in the classical PPM.  

A major difference between PPM$_{cc}$ and PPM appears while decreasing the context length in case of escape symbol
emissions. Shortening the compressed context length by one bit is not appropriate as it causes long chains of escape
sequences. Thus, after emitting an escape, we need to move up in the context trie in steps of a fixed number of bits,
which we refer as \emph{pitch} size throughout the study. 
Aiming to be compatible with the classical solution of decrementing context by one symbol, in PPM$_{cc}$ the pitch size
is assumed to be the \emph{average code length} of the file that is computed during the generation of the
$0^{th}$--order Huffman codes.

\section{Experimental Results}

Experiments are conducted on large Calgary corpus to measure the performance of the CCM--based
implementation PPM$_{cc}$ versus the standard PPM. 

\begin{figure}
  \begin{center}
 \begin{tabular}{c}
 \includegraphics[scale=0.45]{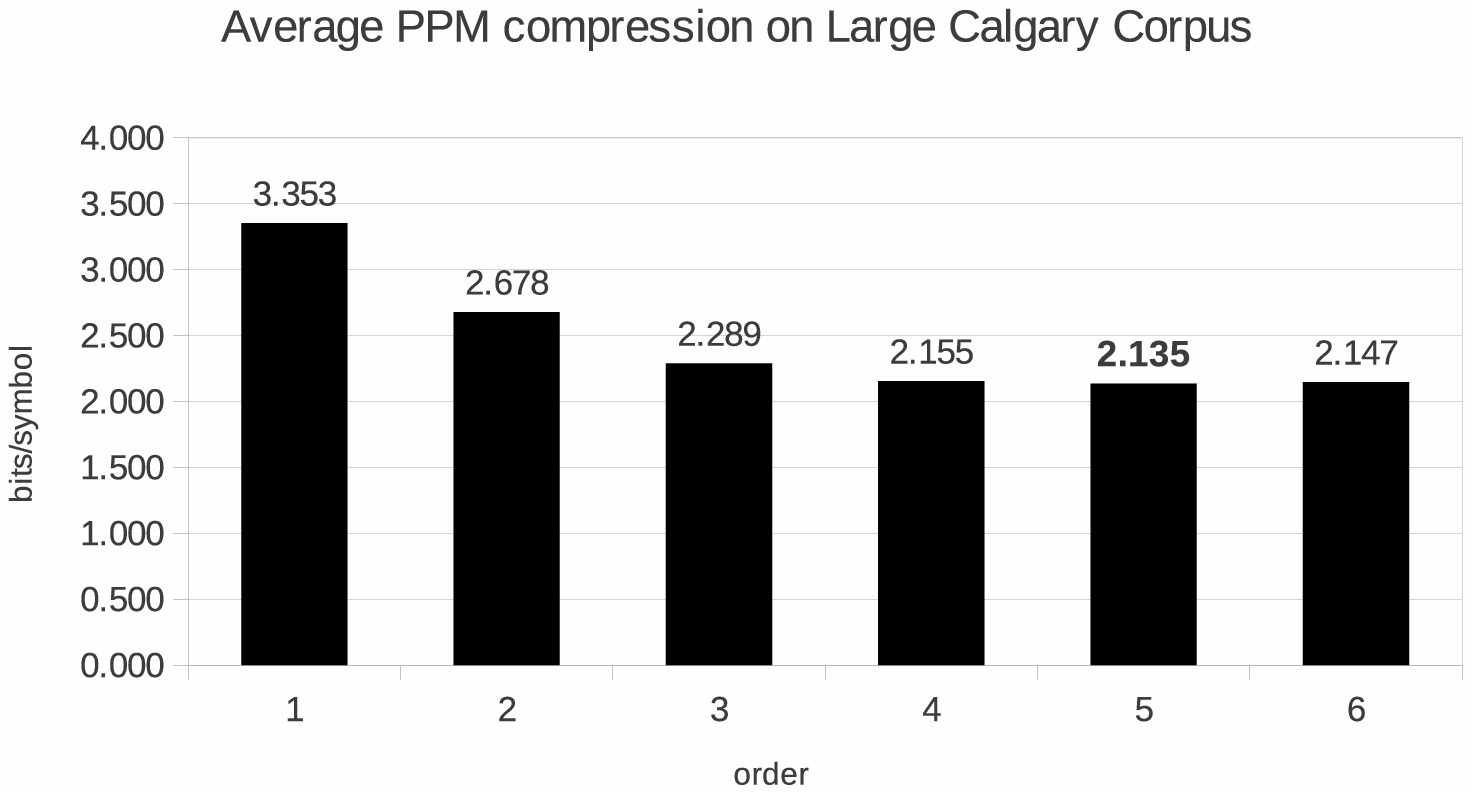}
 % PPMk.eps: 0x0 pixel, 300dpi, 0.00x0.00 cm, bb=0 0 454 255
\\
 \includegraphics[scale=0.45]{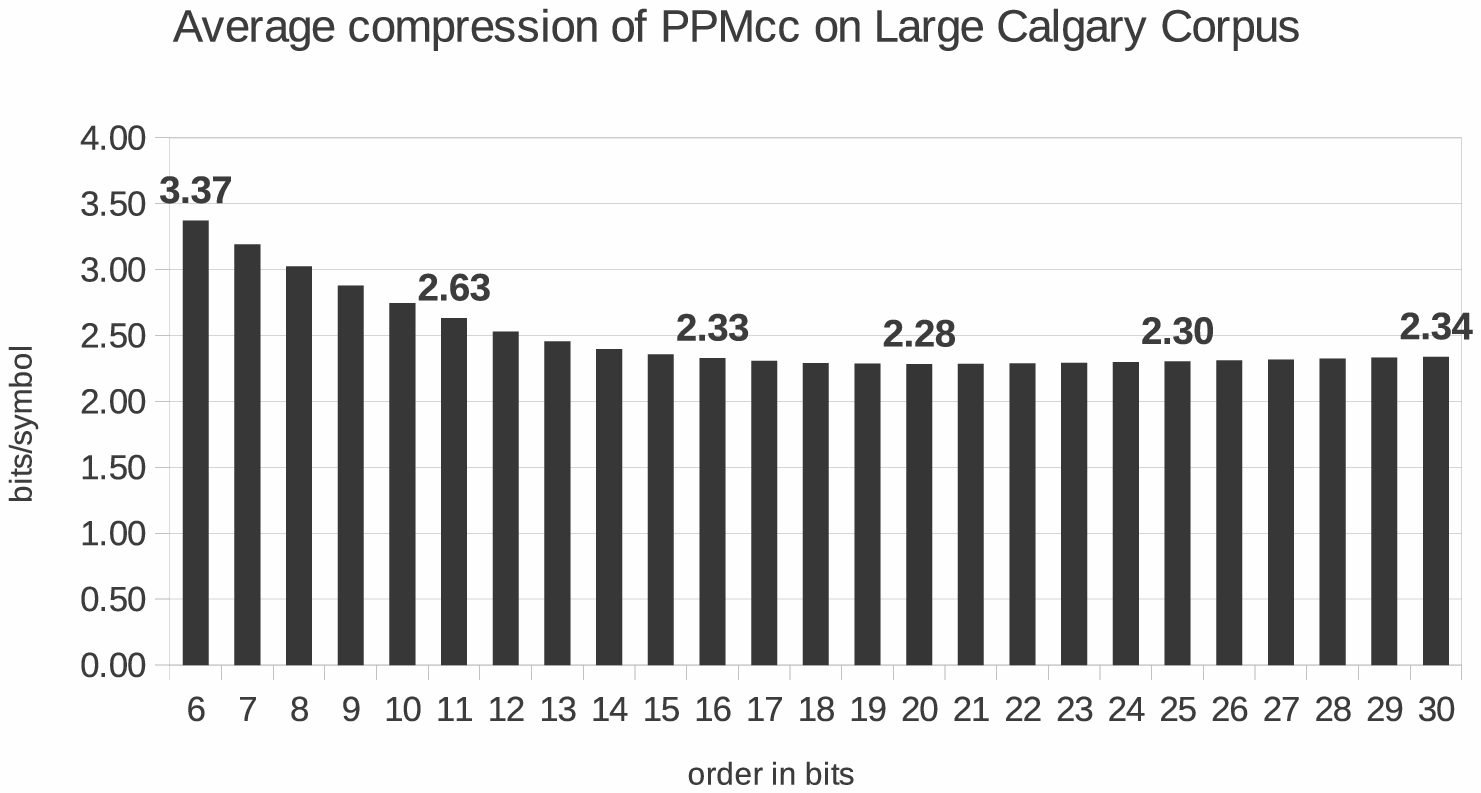}
 % PPMk.eps: 0x0 pixel, 300dpi, 0.00x0.00 cm, bb=0 0 454 255 \\
 \end{tabular}
\end{center}
\caption{The compression ratios of the standard PPM and proposed PPM$_{cc}$ in bits/symbol for different context
lengths.}
\label{fig:compratio}
\end{figure} 

\begin{figure}
  \begin{center}
 \begin{tabular}{c}
 \includegraphics[scale=0.50]{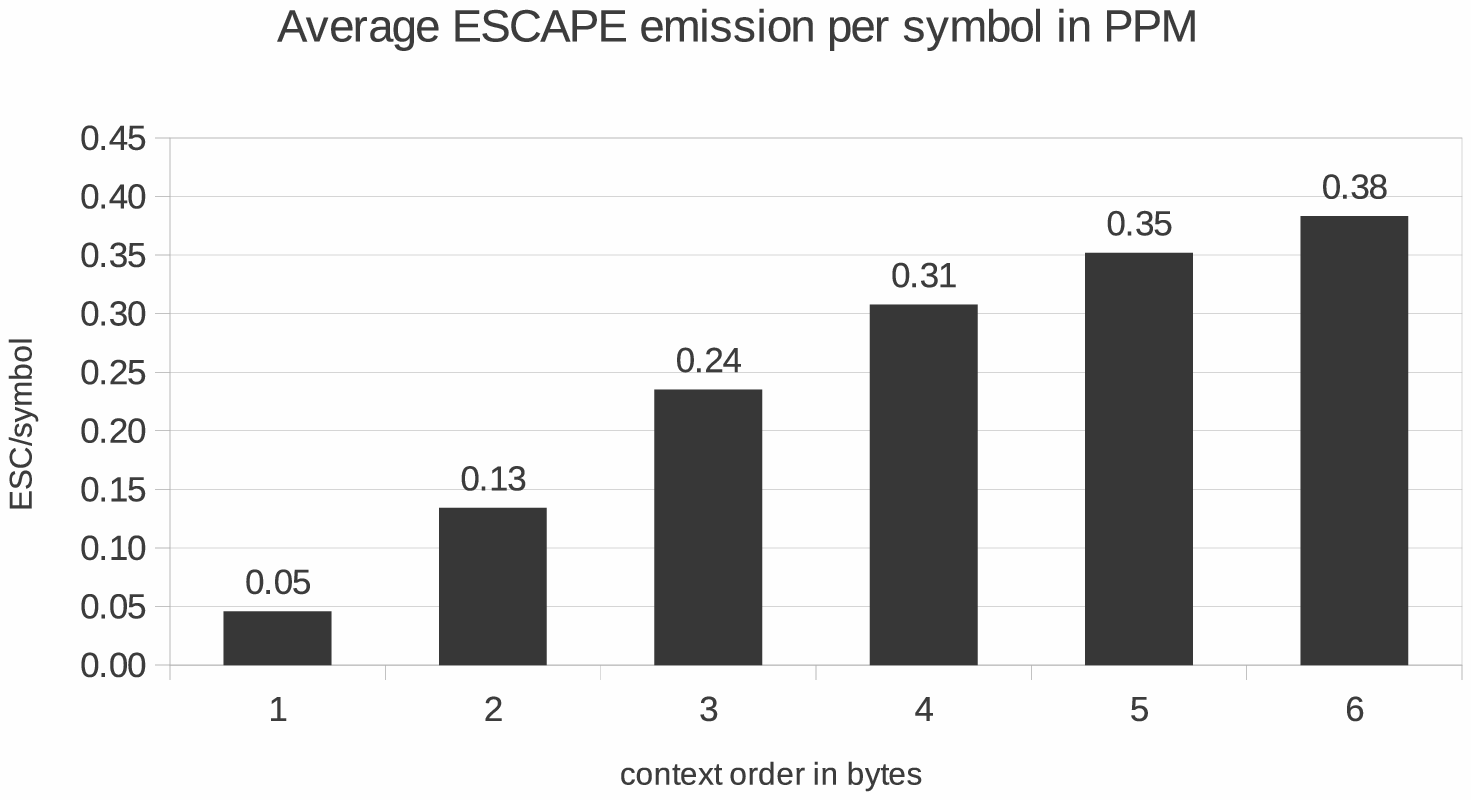}
 % PPMk.eps: 0x0 pixel, 300dpi, 0.00x0.00 cm, bb=0 0 454 255
 \\
 \includegraphics[scale=0.50]{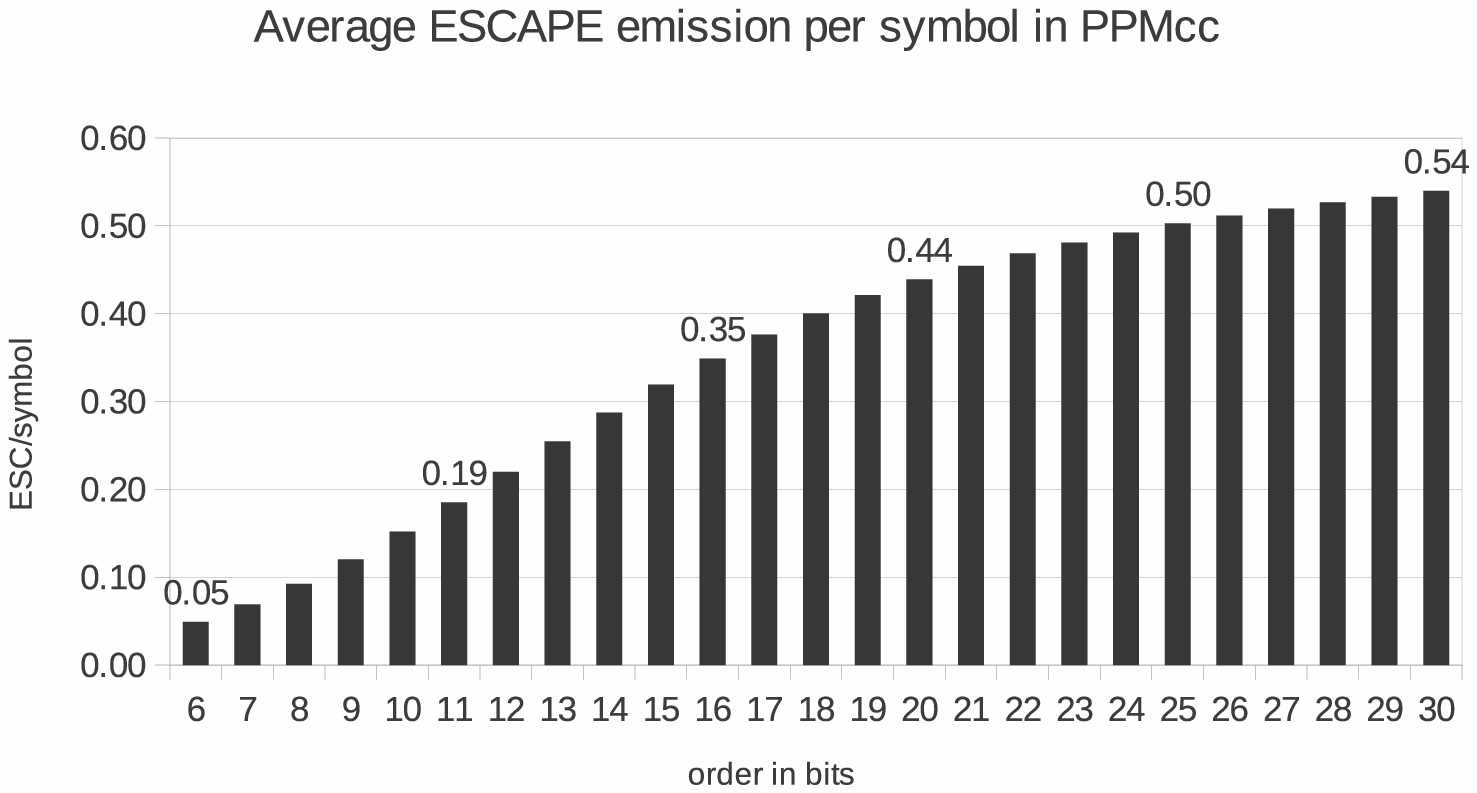}
 % PPMk.eps: 0x0 pixel, 300dpi, 0.00x0.00 cm, bb=0 0 454 255 \\
 \end{tabular}
\end{center}
\caption{Average number of escape characters emitted per symbols in PPM and PPM$_{cc}$.}
\label{fig:esc}
\end{figure}

Figure \ref{fig:compratio} shows the compression ratios achieved by the classical PPM and PPM$_{cc}$. The pitch
sizes  used by PPM$_{cc}$ are $6$ for  \emph{geo}, \emph{obj1}, \emph{obj2}, \emph{trans} files, $2$ for the \emph{pic}
file, and $5$ in all over the rest.
The main concern of this study is to reduce the memory consumption, and thus, trading the compression ratio versus space
is anticipated. 
In general, the competitiveness attained in low orders by PPM$_{cc}$ could not be sustained  in  higher orders due to
the increase in escape character emissions as indicated in Figure \ref{fig:esc} plotting the number of escape characters
emitted per symbol.

The main factor effecting the compression performance is the ambiguity, which is more dominant in high dimensions, 
introduced in CCM. The Huffman codes are uniquely decipherable, and hence, prefix--free, which means the Huffman code
generated for a symbol cannot be a prefix of another one. 
However, the corresponding codes of two distinct characters can share a common prefix. 
When the length of the $\mathcal{C}(t_{i-1}t_{i-2} \ldots t_{i-\ell})$ is longer then $k$, the Huffman code of
$t_\ell$ can only be partially included in the compressed context bit-stream, which might cause an ambiguity. 
As an example, assume that the Huffman codes of letters \texttt{m} and \texttt{n} are $1101100$ and $111010000$
respectively. 
If the last symbol $t_\ell$ in a sample context is \texttt{m} and there is only a two bits vacancy in the compressed
context, then we will append $11$ to the bit-stream, which is also the initials of the code of \texttt{n}. 
In such a case the first $k$--bits of $\mathcal{C}(t_{i-1}t_{i-2} \ldots \mathbf{m})$ and $\mathcal{C}(t_{i-1}t_{i-2}
\ldots \mathbf{n})$ will be equal, and hence, an ambiguity will arise that eventually weakens the statistics.

Another similar problem is caused by the rare characters that have relatively longer Huffman codes. 
Let's assume we are using a $10$--bits compressed context model, where $t_{i-1}$ is \texttt{j} with a $19$ bits long
code. This means we can only use the partial information of \texttt{j} stored in its initial $10$--bits. 
When this $10$ bits long sequence is also a prefix of another symbol's code, the statistics of those contexts are
unified, which in turn causes inefficiency in entropy coding. 

\begin{table}
\begin{center}
\begin{tabular}{c|c|c||c|c|c||c|c|}
\multicolumn{3}{c||}{PPM} &  \multicolumn{3}{c||}{PPM$_{cc}$} & \multicolumn{2}{c}{\% of gain in} \\ \hline
      & Number of &  bits per   &         & Normalized \#            &  bits per &  memory     &
\multicolumn{1}{|c|}{compression} \\
order & Nodes     &  symbol & order   & of Nodes ($\mathcal{Y}$) &   symbol &  space  &   
\multicolumn{1}{|c|}{ratio} \\ \hline 
1 & 83     & 3.603 & 6  & 65      & 3.523 & 21.69 & 2.23 \\
2 & 1909   & 2.907 & 10 & 1017    & 2.860 & 46.73 & 1.60 \\
3 & 15205  & 2.474 & 14 & 10612   & 2.482 & 30.21 & -0.33 \\
4 & 65161  & 2.323 & 18 & 61910   & 2.365 & 4.99  & -1.81 \\
5 & 189280 & 2.325 & 21 & 168448  & 2.369 & 11.01 & -1.91 \\
6 & 417272 & 2.367 & 24 & 369295  & 2.406 & 11.50 & -1.63 \\
\end{tabular}
\end{center}
\caption{Analysis of file \emph{book1} in experiment 1.}
\label{tab:experiment1book1}
\end{table}

The memory usage comparisons of the proposed schemes against the classical {PPM} is achieved by comparing the number of
nodes in the corresponding context tries as this mainly determines the actual space usage. 
A standard node in an ordinary context trie has $|\Sigma|$ integers and $|\Sigma|$ pointers, where there are
$|\Sigma|$ integers and $2$ pointers in the CCM tries as indicated previously in Figure \ref{fig:ContextTrie}. 
Hence, a PPM$_{cc}$ tree with $x$ nodes occupy space equal to an ordinary tree with $\mathcal{Y} = x \cdot \frac{|\Sigma
+ 2|}{2 \cdot |\Sigma|}$ nodes assuming that the integer and pointer types are of same size. 
We refer $\mathcal{Y}$ as the \emph{normalized} number of nodes in CCM trie.

On each file of the corpus we compare order--$I$ PPM$_{cc}$ against order--$k$ PPM, where $I$ is the largest number
that the size of the CCM trie is less than the size of the classic trie. 
A sample analysis performed on file \emph{book1} is given in Table \ref{tab:experiment1book1}. 
The alphabet  of \emph{book1} is of $82$ characters. Thus, while computing the normalized node count
$\mathcal{Y}$, we multiply the number of nodes in the CCM trie with $\frac{82+2}{2\cdot82} = 0.51$. The compressed
context is decremented in orders of $5$ bits that is the rounded average code length $4.56$ bits measured
according to $0^{th}$--order Huffman encoding. The summary of this analysis performed over all files of the Large
Calgary
corpus is given in Table \ref{tab:results}. The best trade=--offs achieved are marked in bold throughout the table.

\renewcommand{\arraystretch}{1.3}
\begin{table}
\small{
\begin{tabular}{l|rr|rr|rr|rr|rr|rr|}
\multicolumn{1}{c}{} & \multicolumn{12}{c}{\% of gain in memory(M) and compression ratio (C)} \\
\multicolumn{1}{c}{} & \multicolumn{2}{c}{order--1} & \multicolumn{2}{c}{order--2} & \multicolumn{2}{c}{order--3} &
               \multicolumn{2}{c}{order--4} & \multicolumn{2}{c}{order--5} & \multicolumn{2}{c}{order--6} \\ 
\multicolumn{1}{c}{} & \multicolumn{1}{c}{M} & \multicolumn{1}{c}{C} &
\multicolumn{1}{c}{M} & \multicolumn{1}{c}{C} & \multicolumn{1}{c}{M} & \multicolumn{1}{c}{C} & \multicolumn{1}{c}{M} &
\multicolumn{1}{c}{C} & \multicolumn{1}{c}{M} & \multicolumn{1}{c}{C} & \multicolumn{1}{c}{M} & \multicolumn{1}{c}{C}\\
\hline
bib & \textbf{23.17} & \textbf{-2.19} & 5.89 & 1.03 & 0.38 & -5.69 & 8.58 & -7.38 & 6.77 & -7.28 & 4.44 & -6.42 \\
\hline
book1 & 21.69 & 2.23 & \textbf{46.73} & \textbf{1.60} & 30.21 & -0.33 & 4.99 & -1.81 & 11.01 & -1.91 & 11.50 & -1.63 \\
\hline
book2 & 34.02 & 0.34 & \textbf{38.45} & \textbf{5.17} & 17.10 & 0.19 & 19.02 & -3.86 & 15.95 & -4.37 & 12.64 & -3.93 \\
\hline
geo & 0.78 & 1.39 & \textbf{28.72} & \textbf{-1.78} & 14.60 & -4.18 & 10.29 & -5.84 & 14.95 & -7.01 & 0.42 & -7.48 \\
\hline
news & \textbf{35.35} & \textbf{-1.24} & 10.86 & 5.09 & 33.35 & -3.83 & 2.88 & -5.65 & 1.32 & -5.85 & 14.47 & -5.41 \\
\hline
obj1 & 1.95 & -4.10 & 4.72 & -11.38 & \textbf{24.00} & \textbf{-13.49} & 7.71 & -14.69 & 11.47 & -14.50 & 1.75 & -14.37
\\ \hline
obj2 & 0.39 & 3.17 & 10.87 & 1.00 & \textbf{19.36} & \textbf{-5.31} & 16.18 & -8.61 & 14.24 & -9.25 & 12.14 & -8.76 \\
\hline
paper1 & 33.33 & -0.80 & \textbf{42.74} & \textbf{-4.99} & 9.58 & -6.89 & 6.96 & -8.12 & 3.26 & -7.53 & 14.86 & -7.26 \\
\hline
paper2 & 30.43 & 0.88 & \textbf{33.94} & \textbf{-0.33} & 31.87 & -5.63 & 23.95 & -5.94 & 16.78 & -5.00 & 11.03 & -4.56
\\ \hline
paper3 & 23.53 & -0.84 & \textbf{28.73} & \textbf{-1.48} & 27.59 & -7.05 & 19.05 & -7.36 & 10.97 & -6.60 & 4.74 & -6.21
\\ \hline
paper4 & \textbf{19.75} & \textbf{-3.11} & 8.37 & -5.92 & 7.23 & -11.29 & 23.31 & -10.95 & 11.08 & -10.54 & 2.39 &
-10.52 \\ \hline
paper5 & \textbf{30.43} & \textbf{-5.19} & 19.88 & -10.01 & 12.29 & -15.19 & 2.26 & -14.19 & 12.68 & -14.16 & 4.43 &
-13.99 \\ \hline
paper6 & 31.91 & -1.98 & 39.55 & -4.9 & \textbf{73.95} & \textbf{-8.62} & 1.91 & -8.93 & 17.03 & -8.67 & 10.76 & -8.27
\\ \hline
pic & 19.38 & 3.55 & \textbf{38.56} & \textbf{1.30} & 9.93 & -3.98 & 24.36 & -6.63 & 21.60 & -9.75 & 14.61 & -10.98 \\
\hline
progc & \textbf{31.18} & \textbf{-2.37} & 5.27 & -3.28 & 11.66 & -9.46 & 7.82 & -9.69 & 3.66 & -9.10 & 0.44 & -9.20 \\
\hline
progl & 27.27 & 2.08 & \textbf{34.11} & \textbf{-3.17} & 1.02 & -6.34 & 0.83 & -7.80 & 17.05 & -7.94 & 10.28 & -8.00 \\
\hline
progp & 28.89 & -0.07 & \textbf{41.22} & \textbf{-8.85} & 10.72 & -10.50 & 9.61 & -9.78 & 6.51 & -9.68 & 3.28 & -9.48 \\
\hline
trans & \textbf{36.00} & \textbf{-3.15} & 23.46 & -4.00 & 29.07 & -14.36 & 1.32 & -9.56 & 1.11 & -9.01 & 0.16 & -8.64 \\
\hline \hline
AVG.  & 23.86 & -0.63 & 25.67 & -2.49 & 20.22 & -7.33 & 10.61 & -8.16 & 10.97 & -8.23 & 7.46 & -8.06 \\ \hline
MAX.  & 35.35 &  3.17 & 46.73  &  5.17  & 73.95 & 0.19  & 24.36 & -1.81 & 21.60 & -1.91 & 14.86 & -14.37 \\ \hline
MIN.  &  0.39 & -5.19 &  4.72  & -11.38 &  1.02 & -15.19 & 0.83 & -14.69 & 1.11 & -14.50 & 0.16 & -1.63 \\ \hline
\multicolumn{13}{c}{} 
\end{tabular}
\caption{PPM$_{cc}$ results.}
\label{tab:results}
}
\end{table} 

\begin{figure}
 \begin{center}
 \includegraphics[scale=0.6]{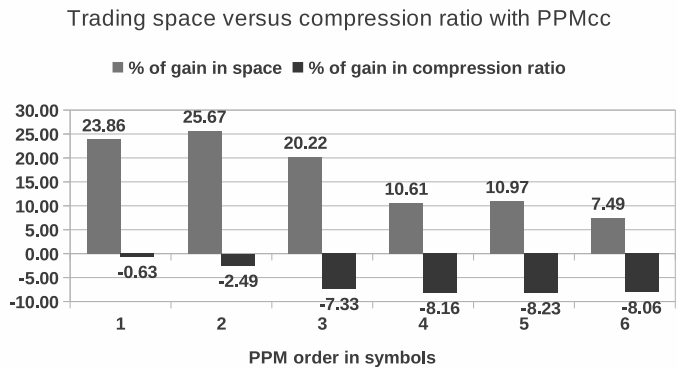}
 % mainresult.eps: 0x0 pixel, 300dpi, 0.00x0.00 cm, bb=0 0 454 255
\end{center}
\caption{The trade--off accomplished between compression ratio and memory consumption.}
\label{fig:mainresult}
\end{figure}

Careful readers should have noticed  that although the individual size of a node is reduced, the number of total nodes
would be much larger when CCM is used as the CCM tree is much less sparse than the tree in the classical context
modeling.  
Thus, the advantage of using small nodes will diminish with the increasing depth of the tree in CCM based
implementation. The experiments complies with this observation that CCM integration is particularly more beneficial in
low orders of PPM. The trade--off between the compression ratio and the memory consumption is depicted in Figure
\ref{fig:mainresult}.

\section{Conclusions}

We have presented a technique to be used in PPM implementations, namely PPM$_{cc}$, based on the
compressed context modeling with the aim of using less space throughout the compression. 
The investigation of the trade--off between the memory consumption and  compression ratio showed that PPM$_{cc}$ is
beneficial especially on low orders, where the gain in space is much more than the sacrifice in compression ratio. The
results depicted in Table \ref{tab:results} reflects interesting observations that the gain in space is much more than
the the loss in the compression ratio. 
It is noteworthy that having a space improvement in low orders is of particular importance as it may help 
the integration of PPM style compressors in mobile environments with less resource requirements. 

Future studies may consider using different encoding techniques while compressing the context such as the higher
order Huffman codes rather than the $0^{th}$ order used in this work. 
Methods to decrease the escape symbol emission rates as well as improving the prediction power with CCM will be
significant with the aim of achieving better compression in less space, where instead of integrating CCM into 
fixed--length models, using the compressed versions of the other context modeling techniques might make sense. 
Studies to investigate CCM usage in other PPM variants may also be interesting. 
%Despite shedding light on reducing the memory consumption in standard PPM, benefiting from compression in compression
%is yet another dimension of future interest. 

\bibliographystyle{splncs03}
\bibliography{ciaa12}

\end{document}